\documentstyle{esub2acm}  

\input epsf
\def\efig#1#2{\hbox{\epsfxsize=#1\epsfbox{#2}}}

\def\Cplusplus{C{\small\raise 0.1em\hbox{$\mathord{+}\mathord{+}$}}}
\newtheorem{theorem}{Theorem}
\newtheorem{lemma}{Lemma}
\def\R{{\bf R}}

\begin{document}
\bibliographystyle{esub2acm}

\title{\Large Fast Hierarchical Clustering and Other Applications of
Dynamic Closest Pairs}

\author{David Eppstein\\
UC Irvine}

\begin{abstract}
We develop data structures for dynamic closest pair problems with
arbitrary distance functions,
that do not necessarily come from any geometric structure on
the objects.
Based on a technique previously used by the author for Euclidean closest
pairs, we show how to insert and delete objects from an
$n$-object set, maintaining the closest pair, in
$O(n\log^2 n)$ time per update and $O(n)$ space. With quadratic space, we
can instead use a quadtree-like structure to achieve an optimal time
bound, $O(n)$ per update.  We apply these data structures to
hierarchical clustering, greedy matching, and TSP
heuristics, and discuss other potential applications in machine learning,
Gr\"obner bases, and local improvement algorithms for
partition and placement problems. Experiments show our new methods to be
faster in practice than previously used heuristics.
\end{abstract}

\category{F.2.2} {Analysis of Algorithms} {Nonnumeric Algorithms} 
\terms{Closest Pair, Agglomerative Clustering}
\keywords{TSP, matching, conga line data structure, quadtree, nearest
neighbor heuristic}

\begin{bottomstuff}
A preliminary version of this paper appeared at the
9th ACM-SIAM Symp. on Discrete Algorithms, San Francisco,
1998, pp. 619-628.
Work supported in part by NSF grant CCR-9258355 and
by matching funds from Xerox Corp.

\begin{authinfo}
\address{Department of Information and Computer
Science, University of California, Irvine, CA 92697-3425,
eppstein@ics.uci.edu}
\end{authinfo}

\end{bottomstuff}

\markboth{D. Eppstein} 
	 {Fast Hierarchical Clustering}
\maketitle

\section{Introduction}

Hierarchical clustering has long been a mainstay of statistical analysis,
and clustering based methods have attracted attention in other fields:
computational biology (reconstruction of evolutionary trees; tree-based
multiple sequence alignment), scientific simulation ($n$-body problems),
theoretical computer science (network design and nearest neighbor
searching) and of course the web (hierarchical indices such as Yahoo).
Many clustering methods have been devised and used in these applications,
but less effort has gone into algorithmic speedups of these methods.

In this paper we identify and demonstrate speedups for a key subroutine
used in several clustering algorithms, that of maintaining closest pairs
in a dynamic set of objects.  We also describe several other applications
or potential applications of the same subroutine, to TSP heuristics, greedy
matching, machine learning, Gr\"obner basis computation, and local
optimization methods.

Although dynamic closest pair data
structures have been studied in low-dimen\-sion\-al geometric spaces
\cite{DobSur-JACM-91,Epp-DCG-95,GolRamSch-SJC-98,Mat-CCCG-93,SchSmiSno-Algo-94,Smi-DCG-92,Sup-SODA-90},
there has been little work on analogous structures in non-geometric spaces,
or in spaces where the dimension is so high as to make taking advantage of
the geometry difficult. However, there are several obvious approaches to
this dynamic closest pair problem. It can be solved by brute force (trivial
recomputation) in time
$O(n^2)$ per update with space $O(n)$, or by a priority queue of distances
in time $O(n\log n)$ per update and space
$O(n^2)$. If we maintain the closest distance itself, and recompute
all distances when we delete one of the two objects forming this
distance, we can even achieve average-case time
$O(n)$ per update, in a model in which any deletion is equally likely.
However, the applications we describe typically repeatedly delete the
closest pair, making the performance of this naive algorithm much worse
than its average case.

Of these naive methods, brute force recomputation may be
most commonly used, due to its low space requirements and ease of
implementation. Three hierarchical clustering codes we examined,
Zupan's \cite{Zup-82}, CLUSTAL~W \cite{ThoHigGib-NAR-94}, and PHYLIP
\cite{PHYLIP} use brute force. (Indeed, they do not even save space by
doing so, since they all store the distance matrix.)  Pazzani's learning
code \cite{Paz-97} also uses brute force (M. Pazzani, personal
communication), as does {\it Mathematica}'s Gr\"obner basis code (D.
Lichtblau, personal communication). The clustering code listed by Anderberg
\cite{And-73} is perhaps more interesting: he uses a ``nearest
neighbor heuristic'' in which one stores the index of each row minimum of
the distance matrix (the nearest neighbor to each point), and only updates
these indices when these minima change. However, this method may
still require $O(n^2)$ time per update in the worst case. Hartigan
\cite{Har-75} describes the same nearest-neighbor heuristic, but resorts to
brute force in the associated code listing.

The purpose of this paper is to show that much better bounds are
possible, using data structures that are simple and likely to be
practical. We adapt and simplify a geometric closest pair data
structure of the author \cite{Epp-DCG-95} to apply in our non-geometric
setting, and show that it achieves nearly the best time and
space bounds above: $O(n\log^2 n)$ time per update and space
$O(n)$.  If linear space is required, this represents an
order-of-magnitude speedup over known solutions.  Further, with quadratic
space, we can also improve significantly on the priority queue; we give
an algorithm based on a quadtree-like structure in the distance matrix,
with time $O(n)$ per update.  This last bound is optimal, since in our
model any algorithm needs to examine all
$n-1$ distances involving each newly inserted object. 
It remains open
whether quadratic space is required to achieve linear time per update.

Along with these theoretical results, we present
experimental results on these data structures and
some simple modifications of them.  In all our experiments, all our new
data structures are preferable to brute force, and one (``FastPair'') is
always preferable to the nearest-neighbor heuristic.  The choice
between it and the other new data structures depends
on problem type and available memory.

For recent geometric applications of similar closest pair data structures,
in problems of dynamic collision detection, offset curve construction, and
skeletonization, see \cite{EppEri-DCG-99}.

\section{Model of Computation}

We assume a model in which we maintain a set of objects subject to
insertions or deletions.  We are also given a bivariate function $d(s,t)$
measuring the distance between objects.  This function need not satisfy
the triangle inequality or other common properties of distances;
indeed, in the Gr\"obner basis application below the distances are not
numbers. We assume only that function values are totally ordered.  The task
of our data structures is to maintain the pair $s,t$ having the minimum
value $d(s,t)$ among all objects in the set.  If two pairs have the same
minimum value, our algorithms may return either pair.

We assume that each object is stored in constant space, that
the distance function can be evaluated in constant time, and that any two
distances can be compared in constant time. These assumptions are not
necessarily valid for all applications; for instance Cheng and Wallace
\cite{CheWal-JAS-93} describe an application of clustering to meteorology,
in which the objects consist of very high dimensional vectors. In
computational biology applications, objects may consist of long sequences
of symbols, and distance evaluations may consist of complicated dynamic
programming routines. In these cases our time bounds can be interpreted as
numbers of evaluations; alternatively, with an additional $O(n^2)$ space, we
can precompute and store the entire distance matrix.

For the clustering applications we describe, we also assume some means
of treating clusters (sets of objects) as objects themselves, and of
computing distances between clusters. There is much freedom in
determining distances between clusters. These distances
need not be the same as the distances between objects, even for clusters
consisting of single objects. Zupan
\cite{Zup-82} describes seven different definitions of distance
between clusters, each of which applies to arbitrary objects and distance
functions, and each of which can be computed in constant time (with
quadratic space to store all cluster distances) by a formula combining
the distances between pairs of subclusters. For biological sequence data,
distances between clusters may be computed by a multiple
sequence alignment that respects previously computed
alignments within each cluster \cite{Cor-NAR-88, Got-CABIOS-94}.
Alternatively, distances may be defined by selecting a cluster member as a
representative object or by combining objects to form a representative in
some application-specific way (e.g., centroids for vector data; consensus
sequences for biological sequence data). The distance between clusters
would then be defined to be the distance between their representative
objects. The multiple fragment heuristic for traveling salesman tours
involves a similar idea in which each cluster is represented by two
objects (at either end of the fragment) with the distance between
clusters equal to the minimum of four distances between representative
objects.

\section{Conga Line Data Structure}

We now describe the dynamic closest pair data
structure from \cite{Epp-DCG-95}, simplified somewhat by
maintaining one set of objects instead of two sets, using a naive
nearest neighbor searching technique in place of geometric range
searching data structures, and relaxing
size restrictions on subsets in a partition of the input.

Our data structure consists of a partition of the dynamic set $S$ into
$k\le\log n$ subsets $S_1$, $S_2$, \ldots, $S_k$, together with a digraph
$G_i$ for each set $S_i$. Each digraph $G_i$ will consist of the union of
a set of directed paths.  Initially all objects will be in $S_1$ and $G_1$
will have
$n-1$ edges. 
$G_i$ may contain edges with neither endpoint in $S_i$; if the number
of edges in all graphs grows to $2n$ we rebuild the data structure by
moving all points to $S_1$ and recomputing $G_1$.
As we will show below,
the closest pair will be represented by an edge in some $G_i$, so we can
find this pair by scanning the edges in all graphs.
As we modify $S$, we create and merge these subsets $S_i$
and associated graphs $G_i$. This involves the following steps:

\begin{description}
\item[Create $G_i$ for a new partition $S_i$.]
When created, $G_i$ will
consist of a single path.  We choose the
first vertex of the path to be any object in $S_i$. Then, we extend the
path one edge at a time. When the last vertex in the partially created path
$P$ belongs to
$S_i$, we choose the next vertex to be its nearest neighbor in
$S\setminus P$, and when the last vertex belongs to $S\setminus S_i$,
we choose the next vertex to be its nearest neighbor in $S_i\setminus P$.
We continue until the path can no longer be extended because $S\setminus P$
or $S_i\setminus P$ is empty.

\item[Merge partitions.]
The update operations described
below can cause $k$ to be too large relative to $n$. If so,
we choose subsets $S_i$ and $S_j$ as close to equal in size as possible: 
more precisely, if $|S_i|\le|S_j|$, we choose these two subsets to
minimize the size ratio
$|S_j|/|S_i|$.  We then merge these two subsets into a single set and create
the graph
$G_i$ for the merged subset as above.
\end{description}

The construction of $G_i$ is essentially the nearest neighbor
TSP heuristic, however we are using it
for a different purpose. The nearest neighbor searches performed when
creating $G_i$ can be done by a naive algorithm that tests all objects in
$S$ or in $S_i$.  Improvements can be made in geometric applications by
applying more sophisticated range search techniques
\cite{Epp-DCG-95,EppEri-DCG-99}. We are now ready to describe the
update operations in this data structure.

\begin{description}
\item[To initialize the data structure]
Create a new subset $S_1$
containing all the initial objects, and create $G_1$.

\item[To insert $x$]
Create a new subset $S_{k+1}=\{x\}$ in the
partition of $S$, create $G_{k+1}$, and merge partitions as necessary
until $k\le\log n$.

\item[To delete $x$]
Create a new subset $S_{k+1}$ consisting of
all objects $y$ such that $(y,x)$ is a directed edge in some $G_i$.
Remove $x$ and all its adjacent edges
from all the graphs $G_i$. Create the graph $G_{k+1}$ for $S_{k+1}$, and
merge partitions as necessary until $k\le\log n$.
\end{description}

\begin{lemma}
The data structure described above correctly maintains the closest pair
in $S$.
\end{lemma}

\begin{proof}
Let $(s,t)$ be a closest pair, where $s$ belongs to a subset $S_i$
created more recently than the subset containing $t$.  Then when $G_i$
was created, it contained $s$, so it contained at least one of $(s,t)$.
Then if $s$ was the first of two vertices added to the path, it must
have chosen as its neighbor either $t$ or a vertex $x$ at least as close
to $s$.  If it chose $t$, edge $(s,t)$ exists in $G_i$.  If it chose
some $x$, then $x$ can not have been deleted, since that would have caused
$s$ to move to a newer $S_j$, so $(s,x)$ is at least as good as
$(s,t)$ and still exists in $G_i$. Similarly if $t$ were chosen first
then it would have formed edge $(t,s)$ in $G_i$ or $(t,x)$ for some
vertex $x$ at least as close to $t$.  Again, $x$ could not have been
deleted because that would cause $t$ to move to a subset $S_j$ created
more recently than $S_i$. So in all cases $G_i$ contains a closest pair.
\end{proof}

\begin{lemma}
The data structure above uses space $O(n)$.
\end{lemma}

\begin{proof}
By construction, the graphs together have at most $2n$ edges (we rebuild
the data structure if this bound is reached), so they take linear space to
store.  The partition is also easily
maintained in linear space.
\end{proof}

\begin{theorem}
The data structure above maintains the closest pair in $S$ in
$O(n)$ space, amortized time $O(n\log n)$ per insertion, and amortized
time $O(n\log^2 n)$ per deletion.
\end{theorem}

\begin{proof}
The correctness and space complexity have already been proven above; it
remains to prove the time bounds. First, let us analyze the time for a
sequence of updates that do not involve rebuilds to the data structure.

We use a potential function argument.
Define the potential of set $S_i$ to be $n|S_i|\log|S_i|$,
and the potential of the whole data structure to be the sum of the
potentials of each subset.
This potential is at most $n^2\log n$ (the
value it would take for a partition consisting of a single set).
The amortized time per operation is $T+B-\Delta$,
where $T$ is the actual time used, $B$ is the increase in the upper
bound $O(n^2\log n)$ on the potential, and $\Delta$ is the increase in
the potential.
Over the course of a sequence of operations, starting
from a situation in which the potential equals $B$, the $B$ and
$\Delta$ terms in this formula telescope, so the total amortized
time for the sequence is at most the total actual time; therefore this
method provides a valid bound on amortized time.

Each time we merge two subsets $S_i$ and $S_j$,
the potential increases by
$$\Delta =
n|S_i|\log{|S_i|+|S_j|\over|S_i|} +
n|S_j|\log{|S_i|+|S_j|\over|S_j|}.$$
Since $|S_i|$ and $|S_j|$ must be within a factor of two of each other,
the two logarithmic terms are constant and this simplifies to
$\Theta(n(|S_i|+|S_j|))$.  Since the path constructed from the merged
subsets has size $O(|S_i|+|S_j|)$, and each edge in the path can be
found in linear time, the total time for the merge is
$O(n(|S_i|+|S_j|))$. Therefore any
time spent performing merges can be
balanced against an increase in the potential function.

Each time we perform an insertion, we create a new set $S_{k+1}$ with
zero potential, and perform $O(n)$ work not counting mergers. However, the
bound $O(n^2\log n)$ on the total potential increases by $O(n\log n)$,
and the amortized time for each insertion must also include this
potential increase, so the total amortized time per insertion is
$O(n\log n)$.

Each time we perform a deletion, we perform $O(n\log n)$ work creating a
new subset of at most $\log n$ objects.  This work is balanced by a
decrease of $O(n\log n)$ in the upper bound on the total potential.
Further, the new set $S_{k+1}$ has some positive potential
(up to $O(n\log n\log\log n)$).
However,
When we move these $\log n$ objects to a new set, the potential
of each set $S_i$ decreases by $\Theta(n\log|S_i|)$ per object,
and this potential decrease dominates the amortized time bound for each
deletion, which is therefore $O(n\log^2 n)$.

To complete this analysis, we estimate the time spent rebuilding the data
structure. Define the {\em excess} of graph $G_i$ to be $|G_i|-2|S_i|$.
Initially, all points are in $S_1$ with a total excess of $-n$.
Each time we merge two subsets, the merged graph's excess becomes
nonpositive.  The only way to create a positive excess is to
move a point out of some $S_i$, by deleting some other point sharing an
edge with the moved point.  Each deletion moves $O(\log n)$ points and
thus increases the total excess by $O(\log n)$.  Therefore, $O(n/\log n)$
deletions need to be performed before each rebuild and the amortized time
per rebuild step is $O(n\log n)$.
\end{proof}

\section{Quadtree Data Structure}

We now describe a simple technique for maintaining the closest pair
even more efficiently, if quadratic space is available.

In a nutshell, we
recursively subdivide the distance matrix into a
quadtree, and maintain the smallest distance within each quadtree square.
Each update affects $O(n)$ squares along a row and column of the distance
matrix, and we update the distances within each square by looking at each
of its four children.

In more detail, assume the objects are numbered $x_0$, $x_1$, $\ldots$,
$x_{n-1}$.  To maintain this numbering, when we insert a new object we
give it the next highest number. When we delete an object $x_i$ we
renumber $x_{n-1}$ to have number $i$, so the numbers stay consecutive.
(In practice, it may be possible to combine a deletion with a subsequent
insertion, and avoid this renumbering step.)

Define subsets consisting of a number of consecutive objects
equal to a power of two:
$S(i,j)=\{x_{i2^j},x_{i2^j+1},\ldots, x_{(i+1)2^j-1}\}$.
Equivalently, let $S(i,0)=\{x_i\}$ and define $S(i,j)$ for $j>0$ to be the
disjoint union of $S(2i,j-1)$ and $S(2i+1,j-1)$.

\begin{lemma}
A set of $n$ objects determines at most $2n-1$ distinct sets $S(i,j)$.
\end{lemma}

\begin{lemma}\label{fewD}
There are $n$ sets $S(i,0)$, each consisting of a single object.
Since each $S(i,j)$ is the disjoint union of two smaller sets,
the number of sets $S(i,j)$ with cardinality $2^j$ is at most half the
number of sets with cardinality $2^{j-1}$, so the total number of sets
is at most $n+n/2+n/4+\cdots=2n-1$.
\end{lemma}

Define $D(i,j,k)$ to be the minimum distance between a point in
$S(i,j)$ and a point in $S(k,j)$.
By the same reasoning as in the
proof of the lemma above, the number of these values is at most $n^2/2 +
n^2/8 + n^2/32 +\cdots=2n^2/3$.

\begin{lemma}
Each insertion or deletion to the set of objects causes $O(n)$
values $D(i,j,k)$ to change.
\end{lemma}

\begin{proof}
We first consider insertion of a new object. This causes $D(i,j,k)$ to
change only when one of $S(i,j)$ or $S(k,j)$ contains the inserted
object. Since for any $j$ there is exactly one $S(k,j)$ containing the
inserted object, the changed values are in one-to-one correspondence
with the sets $S(i,j)$ not containing the inserted object. The result
follows from Lemma~\ref{fewD}.  The argument for deletions is similar,
except that the deletion of one object and renumbering of another causes
roughly twice as many changes to $D(i,j,k)$.
\end{proof}

\begin{theorem}
We can maintain the closest pair among a set of $n$ objects in time
$O(n)$ per insertion or deletion, and $O(n^2)$ space.
\end{theorem}

\begin{proof}
As shown above,
each update causes $O(n)$ changes to the values
$D(i,j,k)$ stored by the data structure. Each changed value can be
recomputed in constant time, using the formula
$$D(i,j,k)=\min \{ D(2i,j-1,2k), D(2i+1,j-1,2k),
D(2i,j-1,2k+1),
D(2i+1,j-1,2k+1)\},$$
if we perform the recomputation for smaller values of $j$ before larger
ones.  The closest pair we seek is $D(0,\lceil\log n\rceil,0)$.
\end{proof}

\section{Hierarchical Clustering Application}

{\em Hierarchical clustering} is the process of forming a maximal
collection of subsets of objects (called clusters), with the property that
any two clusters are either disjoint or nested. Equivalently, it can be
viewed as forming a rooted binary tree having the objects as its leaves;
the clusters then correspond to the leaves of subtrees.
See \cite{And-73, DurOde-74, Har-75, Zup-82} for surveys of the extensive
clustering literature. Although top-down \cite{Yia-SODA-93},
incremental \cite{Zup-82}, and numerical \cite{AgaBafFar-SJC-98}
hierarchical clustering methods are known, hierarchical clustering is
often performed by a bottom up {\em agglomerative} approach. In
agglomerative clustering, one defines a distance between pairs of
clusters based on the distance between objects; then, starting with $n$
single-object clusters, one repeatedly forms new clusters by merging the
closest pair of clusters.

Many variants of agglomerative clustering are known, largely
differing in the definition of cluster distances. This issue was discussed
in more detail in our section on models of computation. For {\em
single-linkage} distance, in which the distance between clusters is formed
by the closest pair of objects, agglomerative clustering reduces to
Kruskal's minimum spanning tree algorithm, and can be performed in $O(n^2)$
time and $O(n)$ space by instead applying Prim's or Boruvka's algorithm and
sorting the MST edges.  There has been some recent work on clustering in
low-dimensional spaces \cite{KrzLev-DCG-98} or with Hamming distances on
binary data \cite{Aic-PhD-97}.
But for cluster distances other than single linkage in more general data
sets, no such speedups are known to the
merging process defined above. Our data structures improve these clustering
algorithms by allowing the nearest pair of clusters to be found quickly.

\begin{theorem}\label{cluster}
We can perform bottom-up hierarchical clustering, for any cluster
distance function computable in constant time from the distances between
subclusters, in total time $O(n^2)$.  We can perform median, centroid,
Ward's, or other bottom-up clustering methods in which clusters are
represented by objects, in time $O(n^2\log^2 n)$ and space $O(n)$.
\end{theorem}

\begin{proof}
Each step in these clustering algorithms can be performed by finding the
closest pair of clusters, deleting these two clusters from the set of
objects represented by our closest pair data structure, and inserting a
new object representing the new merged cluster.
\end{proof}

\section{Traveling Salesman Heuristic Application}

Since the traveling salesman problem is NP-complete, but has many
applications, a number of heuristics have been devised to approximately
solve it.  Some, such as the nearest neighbor
heuristic (discussed above in connection with our low-space closest
pair data structure) and the double minimum spanning tree heuristic,
can be solved easily in quadratic time and linear space (optimal in our
non-geometric model of computation).  However, Bentley \cite{Ben-SODA-90}
has shown that these simple techniques are outperformed by other,
seemingly harder to compute methods, such as the {\em multiple fragment
heuristic}: consider all edges one at a time in sorted order,
and include an edge if it connects the endpoints of two {\em fragments} of
tours (connected components of previously added edges).

\begin{theorem}
We can implement the multiple-fragment heuristic in time $O(n^2)$
or in time $O(n^2\log^2 n)$ and space $O(n)$.
\end{theorem}

\begin{proof}
This can be seen as a type of hierarchical clustering, in which
clusters correspond to fragments, and the distance between two clusters
is the length of the shortest edge connecting their endpoints.
The sequence of edges added by the hierarchical clustering algorithm of
Theorem \ref{cluster} is then exactly the same as the sequence added in
the multiple fragment method.

Alternatively, instead of maintaining the closest pair among a set of
clusters, maintain the set of fragment
endpoints, with distance $+\infty$ between endpoints of the same
fragment. Each step of the algorithm then consists of selecting the closest
pair, deleting one or both of these endpoints (if they belong to nontrivial
fragments) and modifying the distance between the endpoints of the
combined fragment.
\end{proof}

Another TSP heuristic, {\em cheapest insertion} \cite{RosSteLew-SJC-77},
maintains a tour of a subset of the sites, and at each step adds
a site by replacing an edge of the tour by two edges
through the new site. Each successive insertion is chosen as the one
causing the least additional length in the augmented tour.

\begin{theorem}
We can implement the cheapest insertion heuristic in time $O(n^2)$
or in time $O(n^2\log^2 n)$ and space $O(n)$.
\end{theorem}

\begin{proof}
We use our data structures to maintain a set of $n$ objects: the $k$ edges
in the tour after the $k$th insertion, and the $n-k$ remaining uninserted
sites.  The distance between an edge and a site is defined to be the
increase in length that would be caused by the corresponding insertion;
all other distances are~$+\infty$. In this way each successive insertion
can be found as the closest pair in this set.
\end{proof}

For sites in a vector space or other set for which the distance
between sites and edges is well defined, we can similarly
implement {\em nearest insertion} \cite{RosSteLew-SJC-77},
which inserts the object closest to the current tour.  How efficiently we
can implement the farthest insertion heuristic remains unclear.

\section{Greedy Matching Application}

The {\em greedy matching} of a set of points, with some distance
function, is found by repeatedly selecting and removing the pair of
points with minimum or maximum distance, depending on whether one wants
a minimum- or maximum-weight matching.  This technique was introduced by
Reingold and Tarjan \cite{ReiTar-SJC-81}, who noted that greedy matchings
could be constructed in $O(n^2\log n)$ time by sorting the set of
distances. Since that paper there has been no improvement in the time
bounds for greedy matching.

Greedy matching is not a particularly good approximation to
the minimum weight matching \cite{ReiTar-SJC-81}, even in the average
case for one-dimensional points \cite{FriMcDRee-SJC-90}.  However, for
maximum weight matching with non-negative inter-object distances, greedy
matching comes within a factor of two of optimal, and may provide a good
starting point for augmenting-path based techniques for finding optimal
matchings.

Greedy matching may also be appropriate for non-numeric distances for
which addition is undefined, since it lexicographically minimizes or
maximizes the set of edge weights in the matching.

\begin{theorem}
We can perform greedy matching in time $O(n^2\log^2 n)$ and space
$O(n)$, or in time $O(n^2)$.
\end{theorem}

\begin{proof}
We use the data structures defined above to
repeatedly find and delete the closest pair.
\end{proof}

\section{Other Applications}

We now discuss some other potential applications of our data structures,
in which the savings they achieve are less easy to quantify.

\subsection{Gr\"obner Bases}

We first consider the problem of computing Gr\"obner bases for
polynomial ideals.  Buchberger's Gr\"obner basis algorithm is a key
component of many symbolic algebra systems and has found a large number of
applications including computational geometry and robotics
\cite{Buc-TCA-87}, automated deduction \cite{CleEdmImp-STOC-96}, and
combinatorial enumeration \cite{Stu-96}.  This algorithm takes as input a
set
$B=\{f_1,f_2,\ldots\}$ of polynomials and a {\em term ordering} for
comparing monomials, and proceeds to modify $B$ in a sequence of steps,
in which {\em $S$-polynomials} $S(f_i,f_j)$ are constructed and added to
$B$, and polynomials in $B$ are simplified by
subtracting multiples of each other. As the algorithm proceeds, $B$ can
grow very large, so space efficiency is crucial. Further, the choice of
which $S$-polynomial to form can make a large difference in the
algorithm's efficiency. For this reason, many
implementations follow a suggestion of Buchberger to use the {\em normal
selection strategy} (e.g. see \cite[p.~130]{AdaLou-94}): select $f_i$ and
$f_j$ for which the least common multiple of the leading terms of
$f_i$ and $f_j$ is as small as possible in the term ordering. (Other
selection strategies have also been proposed
\cite{Cza-ISSAC-91, GioMorNie-ISSAC-91} and it seems likely that our
methods apply as well to them.)

We can easily apply our closest pair data structures to maintain the set
$B$ and select the appropriate pair $f_i$, $f_j$.  Distances between
members of $B$ can be measured by least common multiples of leading
terms; these values, although non-numeric, can be compared by
the term ordering.  One complication arises, however: once we have
processed $S(f_i,f_j)$, we do not want to select the same pair again. So,
some data structure such as a hash table should be used so we
can test whether this $S(f_i,f_j)$ has already been computed,
and if so modify the distance between $f_i$ and $f_j$ to $+\infty$.  Such a
modification can performed as efficiently as an insertion in our
linear-space data structure: simply move $f_i$ and $f_j$ to a new subset in
the partition of the objects maintained by the data structure. In our
quadtree data structure, no hash table is needed and modification of a
single distance is even easier, taking time
$O(\log n)$.

However, pair selection forms a small part of the runtime of
Buchberger's algorithm (D. Lichtblau, personal communication) so
improvements would likely have to be made elsewhere to make it worthwhile to
implement our data structures for this application.

It may also be of interest to consider applying our techniques to other
pair-combination methods of automated deduction such as resolution-based
theorem proving.

\subsection{Constructive Induction}

A second potential application arises in machine learning.  Constructive
induction
is a technique for synthesizing new attributes for multi-attribute data, by
combining pairs of attributes. This method can be used to enhance
learning methods that can not represent such combinations directly, or that
are based on an assumption of attribute independence
that may not hold in the actual input.  For example, Pazzani \cite{Paz-97}
forms new attributes from Cartesian products of pairs of discrete-valued
attributes, and demonstrates improvements to the learning abilities of
Bayesian and nearest-neighbor classification systems. In Pazzani's
experiments, each new product attribute is chosen greedily, as the one that
leads to the biggest improvement as measured by leave-one-out
cross-validation. Such greedy pairwise combination again seems a natural
application for our data structures, but we can only apply them if the
quality of an attribute combination stays stable
while we insert or delete unrelated attributes. According to Pazzani
(personal communication), this stability does hold in practice.

\subsection{Non-Hierarchical Clustering}

Duran and Odell \cite[p.~23]{DurOde-74} describe a non-hierarchical
clustering procedure due to Ball and Hall \cite{BalHal-TR-65}, to which
our methods may also apply. In this procedure, a clustering is improved
by repeatedly merging the closest pair of clusters (measured by average
squared distance) and splitting the cluster with the highest variance.
Clearly, our data structures can be used for the merge step, but it is
not clear whether this is a significant part of the overall
complexity of the algorithm, which also includes ``$K$-means''-like phases
in which clusters are reconstructed by moving objects to the
nearest cluster centroid.

\subsection{Local Optimization}

Local search procedures such as two-optimization are a common method for
improvement of heuristic solutions to optimization
problems such as parts placement, traveling salesman tours, or graph
partitioning.  In these procedures, one modifies a suboptimal solution by
moving a small number of objects; the ``two-'' in two-optimization refers
to the number of objects moved.  So, for instance, in graph partitioning,
one maintains a correct partition while improving the number of
crossing edges, by swapping one vertex on one side of the
partition for a vertex on the other side; in the traveling salesman
problem, one maintains a correct tour while improving its
length by removing two edges and replacing them by two other edges
connecting the same four vertices. Our
methods can likely be used in some of these problems, to maintain the pair
of objects the replacement of which leads to the greatest improvement in the
objective function.

However, in practice, local optimization is often
combined with techniques such as simulated annealing, which
randomly selects local changes and allows the objective function
to become worse in an attempt to escape local minima.  It is not clear how
techniques for maintaining the best local improvement should be combined
with this simulated annealing approach.
Further, application of our ideas to e.g. TSP two-optimization is
complicated by the fact that only one of the two ways of replacing a pair
of edges will lead to another valid tour; it is not clear whether our data
structure can be modified to deal with this additional complication, or
with similar complications arising in other problems.

\section{Implementation and Experiments}

\subsection{Algorithms Implemented}

In order to test our data structures, we implemented them in a testbed
of four algorithms: greedy matching, the multi-fragment TSP
heuristic, the cheapest insertion TSP heuristic, and hierarchical
clustering by unweighted medians (UPGMA).

We implemented several
methods for generating random objects: uniformly
distributed vectors with various distance functions (including dot product
as well as the more familiar $L_1$, $L_2$, and $L_\infty$ metrics),
hierarchically clustered points (via a generalization of the Sierpinski
tetrahedron fractal), random leaves of a large binary tree, and random
distance matrices.  Each object generator allowed all distances to be
negated, forming a maximization rather than minimization problem.

The data structures we implemented included our own conga line and
quadtree methods, brute force , and the
``nearest neighbor heuristic'' suggested by authors including Anderberg
\cite{And-73}.  In this method, we store each point's
nearest neighbor; closest pairs can be found by scanning this
list of neighbors.  Insertions can be performed in $O(n)$ time by computing
the nearest neighbor to the inserted point and testing whether it should
become the nearest neighbor of any other point.  However, when a point is
deleted, any other point for which it is nearest neighbor must find a
new neighbor; if the deleted point was neighbor to $k$ other points, the
neighbor heuristic takes time $O(kn)$.  If deletions are random or the
points belong to a low dimensional metric space, $k=O(1)$ and the time per
update is $O(n)$, but it is not hard to find examples in which the worst
case time per update is $\Theta(n^2)$.
We did not implement the priority
queue method due to its complexity, high space usage and expected
poor performance.

In all the methods we implemented, nearest neighbors were
computed by a naive sequential scan through all points.
In many applications, nearest neighbors can be computed more quickly by
heuristics such as spiral search; however we did not implement this due to
its complexity.  We believe that faster searching would equally speed up
brute force, the neighbor heuristic, and our conga line based methods,
so adding such heuristics should not change our overall experimental
conclusions except possibly by making the quadtree method (which can not
use fast neighbor-finding methods) appear worse.

\begin{figure}[t]
$$\efig{3.25in}{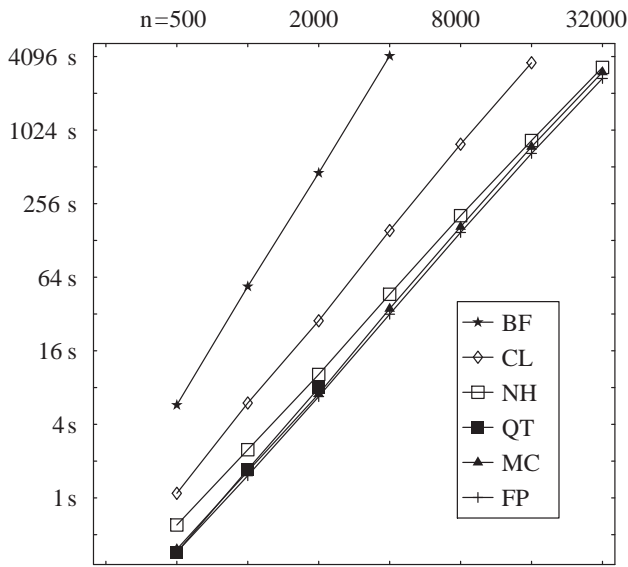}$$
\caption{Hierarchical clustering in $\R^{20}$. Points
were placed  uniformly at random in the unit hypercube and clustered by
unweighted medians.}
\end{figure}

\begin{figure}[t]
$$\efig{3.25in}{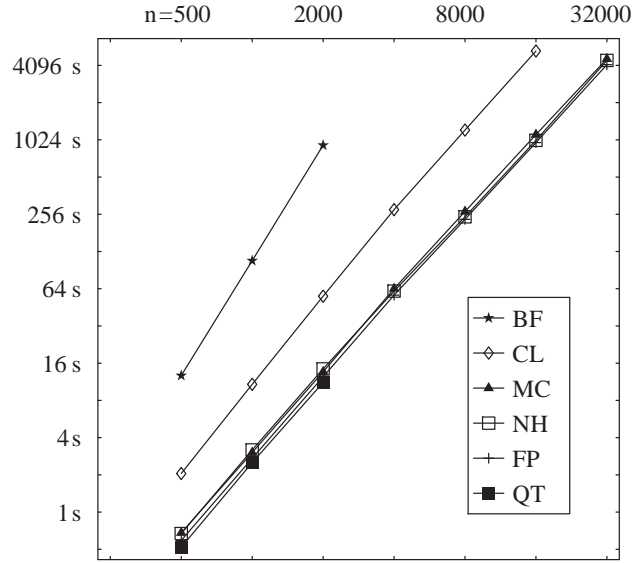}$$
\caption{Hierarchical clustering in a 31-dimensional fractal. Points
were placed  uniformly at random in a generalized
Sierpinski tetrahedron formed by choosing 5 random binary values and
taking bitwise  exclusive ors of each nonempty subset, and clustered by
unweighted medians.}
\end{figure}

\begin{figure}[t]
$$\efig{3.25in}{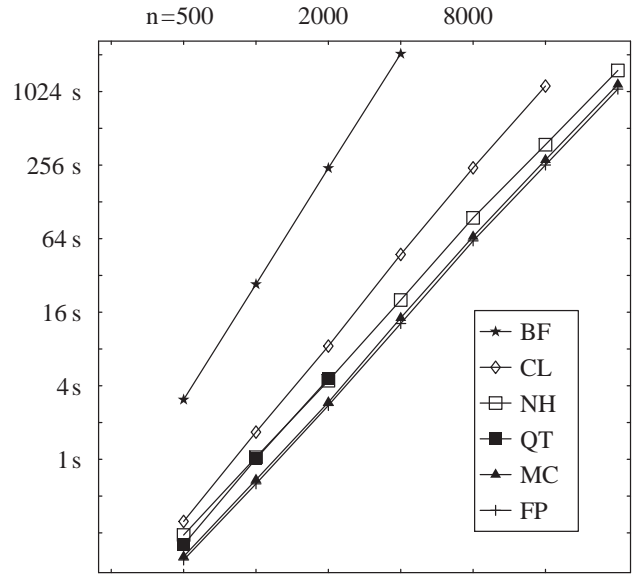}$$
\caption{Greedy matching of points
placed  uniformly at random in the unit hypercube in $\R^{20}$.}
\end{figure}

\begin{figure}[t]
$$\efig{3.25in}{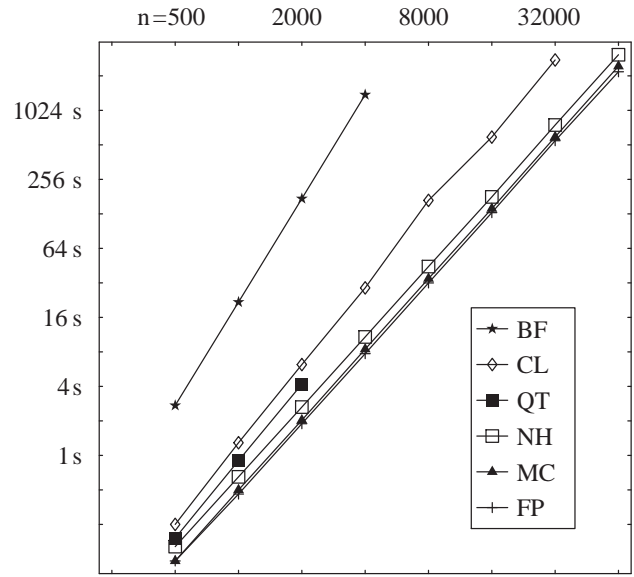}$$
\caption{Greedy matching of points
with pseudorandom distances. The distance between two points was computed
by using their indices to modify the seed for the drand48 random number
generator.}
\end{figure}

\begin{figure}[t]
$$\efig{3.25in}{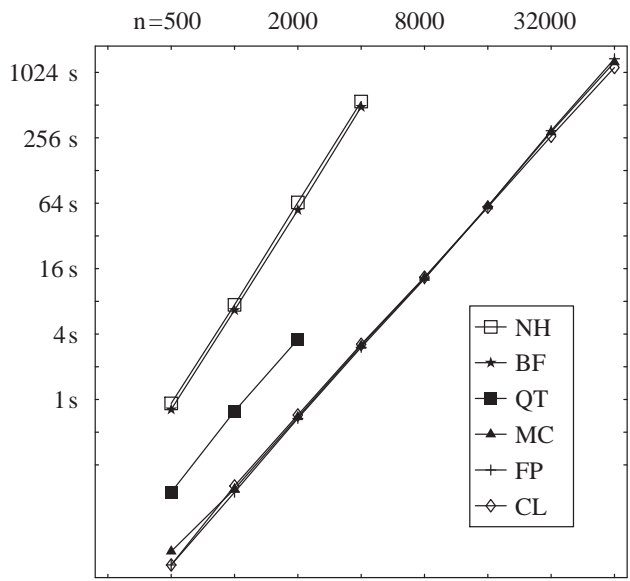}$$
\caption{Greedy maximum-weight matching of points placed uniformly at
random in the unit square, with the $L_1$ metric.}
\end{figure}

\begin{figure}[t]
$$\efig{3.25in}{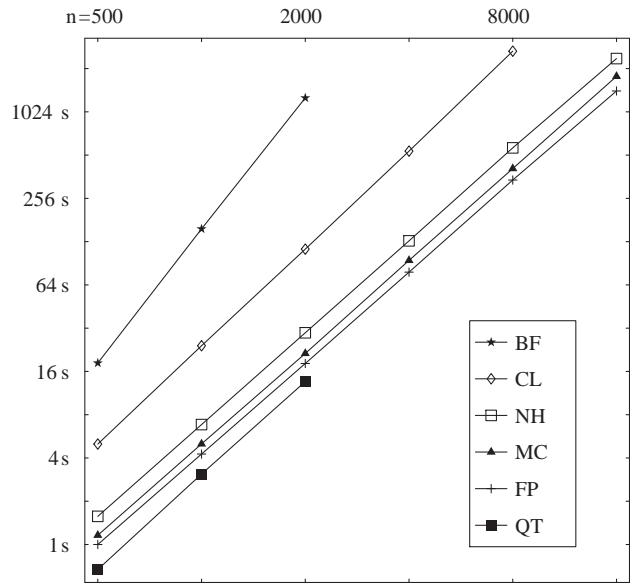}$$
\caption{Cheapest insertion heuristic for TSP of points placed uniformly
at random in a 20-dimensional hypercube.}
\end{figure}

\begin{figure}[t]
$$\efig{3.25in}{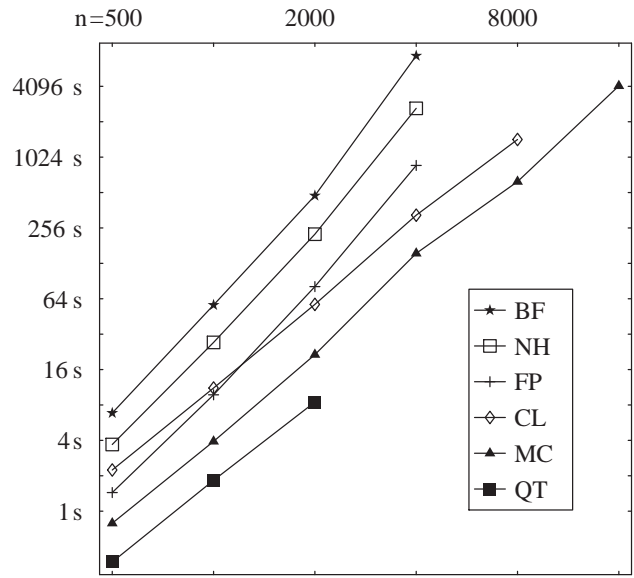}$$
\caption{Most expensive insertion heuristic for MAXTSP of points placed
uniformly at random in the unit square, with the $L_1$ metric.}
\end{figure}

\begin{figure}[t]
$$\efig{3.25in}{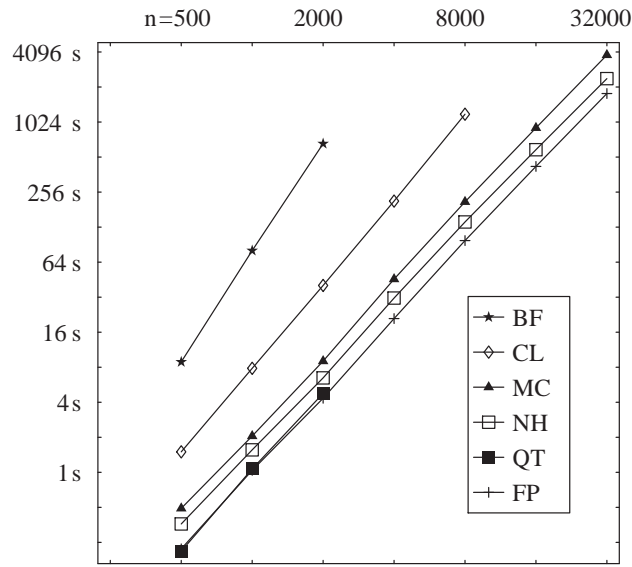}$$
\caption{Multifragment heuristic for TSP of points
placed  uniformly at random in the unit hypercube in $\R^{20}$.}
\end{figure}

\begin{figure}[t]
$$\efig{3.25in}{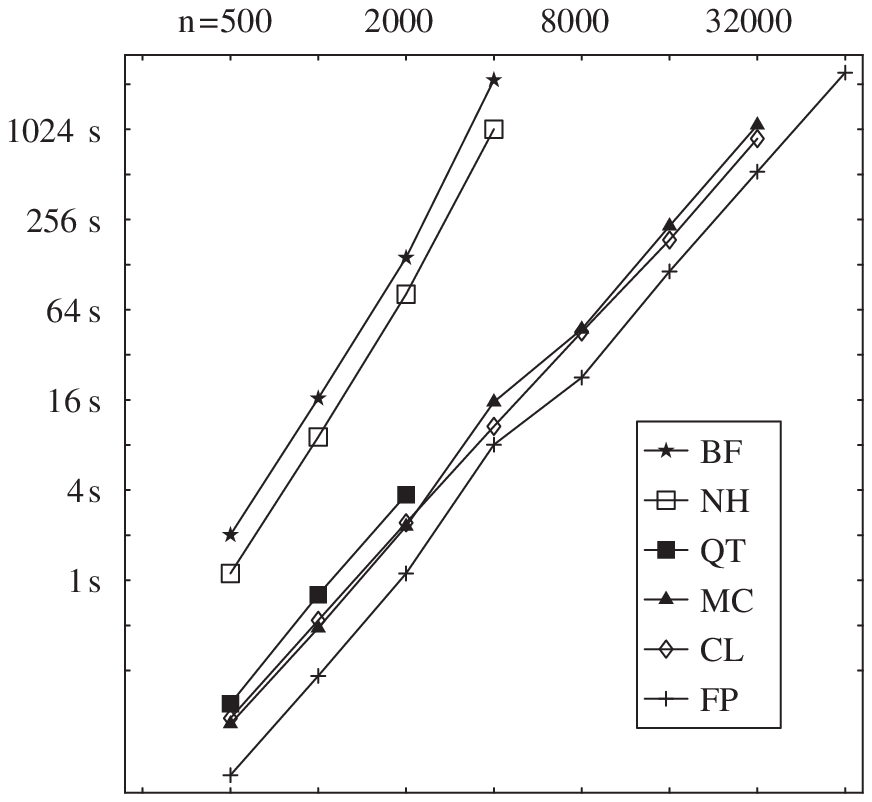}$$
\caption{Multifragment heuristic for MAXTSP of points placed
uniformly at random in the unit square, with the $L_1$ metric.}
\end{figure}

\subsection{Simplified Data Structures: MultiConga}

Our conga line implementation includes a
parameter $k$ for the number of subsets into which to partition the objects.
For best results in our theoretical analysis,
$k$ should be $\Theta(\log n)$; our implementation's default
is $k=\log_2 n$.  Our initial expectation was that multiplying this default
by a small constant might lead to small improvements, but that
non-logarithmic values would cause the time to blow up. To our surprise,
the data structure became faster as $k$ grew very large, until the number of
distance computations stabilized but the overhead of maintaining many
subsets slowed down the structure.

In retrospect, we can explain this heuristically: if $k$ is large, we do
few merges of existing subsets, reducing the amortized time per insertion.
Although the worst case number of points moved to a new subset by each
deletion is $k$, the expected number (if any deletion is equally likely)
is $O(1)$ regardless of the number of subsets, so
increasing $k$ could typically cause less harm than our worst case analysis
would suggest.

With this experience and heuristic justification, we decided to try a
modified version of the conga line structure, which we call the
``multiple-subset conga line'' or ``MultiConga'' for short.  In this
structure, we simply never merge subsets $S_i$; instead, whenever an
insertion or deletion creates a new subset we let the total number of
subsets grow.  More formally, we modify the conga line data structure
operations as follows.

\begin{description}
\item[To initialize the data structure]
Create a new subset $S_1$
containing all the initial objects, and create $G_1$.

\item[To insert $x$]
Create a new subset $S_i=\{x\}$ in the
partition of $S$, and create $G_i$ (consisting of a single edge from
$x$ to its nearest neighbor).

\item[To delete $x$]
Create a new subset $S_i$ consisting of
all objects $y$ such that $(y,x)$ is a directed edge in some $G_j$.
Remove $x$ and all its adjacent edges
from all the graphs $G_j$. Create the graph $G_i$ for $S_i$.
\end{description}

In our experiments, this variant of our closest pair data structure was
usually faster than the original conga line data structure, sometimes much
faster than the neighbor heuristic, and only rarely slightly slower than the
neighbor heuristic.  We can provide theoretical evidence for its speed:

\begin{theorem}
The MultiConga method described above correctly maintains the closest pair
in amortized time $O(n)$ per insertion and $O(n^{3/2})$ per deletion.
\end{theorem}

\begin{proof}
Correctness follows from the correctness of the conga line data
structure. To prove the time bound, we use a potential function
$\varphi = \sum_i |S_i|^2 n^{1/2}.$
Each insertion changes this potential by $n^{1/2}$ and takes time $O(n)$.
Each deletion in which $k$ points are moved to a new subset takes time
$O(kn)$, but increases $\varphi$ by
$k^2n^{1/2} - O(n^{3/2})$.  For any $k$,
the amortized time (actual time minus difference in $\varphi$) is
$O(kn - k^2n^{1/2} + n^{3/2})=O(n^{3/2})$.
\end{proof}

Although one can concoct examples in which this worst-case bound is tight,
we did not find any natural problem for which this method achieved its
worst case.

\subsection{Simplified Data Structures: FastPair}

Since the expected number of points moved into a cluster on each deletion
is small, we decided to try a further simplification.
In the ``FastPair'' method, like MultiConga, we never merge
subsets.  But further, in the case that a deletion would cause $k$ points
to move from their current subsets to a new subset, we instead form $k$
singleton subsets.  More formally, we have the following two operations:

\begin{description}
\item[To initialize the data structure]
Create a new subset $S_1$
containing all the initial objects, and create $G_1$.

\item[To insert $x$]
Create a new subset $S_i=\{x\}$ in the
partition of $S$, and create $G_i$ (consisting of a single edge from
$x$ to its nearest neighbor).

\item[To delete $x$]
Create a separate new subset $S_i=\{y\}$ for each
object $y$ such that $(y,x)$ is a directed edge in some $G_j$. Remove $x$
and all its adjacent edges from all the graphs $G_j$. Create the graph
$G_i$ for each newly created subset
$S_i$.
\end{description}

The advantage of this data structure compared to the previous ones is that
each object $x$ has an outgoing edge to a neighbor only within the set to
which it belongs.  Therefore, the actual data stored in the structure need
only consist of the weight of
this outgoing edge and the identity of the neighbor it points to.
The partition of the objets into subsets $S_i$ does not need to be stored
explicitly, as it is not used by the update operations.
In addition, the number of edges in the structure is always at most $n$,
so we need not worry about rebuilding when the number of edges grows too
large.

This FastPair data structure closely resembles the nearest neighbor
heuristic, in which again each point remembers a single neighbor.
However in the FastPair heuristic the stored neighbor may not always be
nearest.  In the initial construction of the data structure, instead of
computing nearest neighbors for each point, we construct a single conga
line, in order to maintain some control over the number of incoming edges
per object.  And, when inserting a new point, we compute its nearest
neighbor as before, but we do not change the stored neighbors of other
points even if the newly inserted point is nearer than these stored
neighbors. Like the nearest neighbor heuristic, the FastPair method takes
linear expected time for random deletions, but has a quadratic worst case.
In our experiments, FastPair was always faster than the neighbor heuristic.

\subsection{Experimental Results}

Log-log charts of timing results from our computational experiments are
presented in Figures 1--9.  In the figures, ``BF'' stands for the brute
force method, ``NH'' for the neighbor heuristic, ``QT'' for our quadtree
method, ``CL'' for the basic conga line method, ``MC'' for MultiConga, and
``FP'' for FastPair.
 The times include only the construction of the
closest pair data structure and algorithm execution (not initial point
placement) and are averages over ten runs.
The algorithms were implemented in {\Cplusplus},
compiled and optimized by Metrowerks Codewarrior 10, and run on a 200MHz
PowerPC 603e processor (Apple Powerbook 3400c).
The quadtree data structure was limited to 1000 points by its high memory
requirements; other data structures were tested up to the point where a
larger input would not fit comfortably into an overnight test run.

We ran one representative application (greedy matching)
using a variety of distance functions, and ran a selection of other
applications on two distance functions for which our
data structures exhibited strikingly different qualitative
behavior (Euclidean closest pairs for uniformly generated points in
$\R^{20}$, and rectilinear farthest pairs for uniformly generated points in
the unit square).  For hierarchical clustering, we also
ran a further test on a point set with a fractal structure, to
test whether the behavior we observed on uniform points could be assumed
to hold also for more realistic clustered data.

Each application performed linearly many
updates, so linear time per update translates to quadratic total time in our
tests, and quadratic time per update translates to cubic total time.
Asymptotic runtime can be estimated by examining the change in running time
when doubling the problem size; if the time increases by a factor
of four, it can be estimated as quadratic or nearly quadratic,
while if the time increases by a factor of eight, it can
be estimated as cubic.  Due to caching and other issues, it was common for
times to increase by factors larger than the theoretical worst
case bound, but in general all experiments gave results
consistent with cubic or quadratic runtimes.

As expected, brute force always gave
cubic runtimes.  The neighbor heuristic was often quadratic, but
on some problems was cubic, even sometimes slower than brute
force.  FastPair was also sometimes quadratic, and sometimes
cubic; however it was the only method to consistently run faster than the
neighbor heuristic (sometimes by a linear factor).
The remaining methods always exhibited quadratic behavior
(although MultiConga could theoretically have a slower worst
case) but sometimes differed by factors of three or more in total runtime. 
The quadtree method was surprisingly slow; although it
performed few distance computations, it was
generally only faster than other methods for problems with expensive
distance computations. The basic conga line method was
often slower by a factor of three to five than its simplifications, and on
some problems this factor seemed to be increasing with $n$, perhaps showing
that the logarithmic factors in its theoretical time bound were
active in practice.

Our conclusion would be to use the quadtree method for problems with few
points and slow distance computations; to use FastPair for most
applications (after testing to verify that it behaves well for the given
application) and to use MultiConga or occasionally the original Conga
Line structure when FastPair is known to behave poorly
or when a more robustly fast method is required.

The problem of caching remains interesting.  The methods we tested
involve sequential scans through memory, a behavior known to
reduce the effectiveness of cached memory.  Some effects of this
appear in our data; for instance the last two rows of the brute force
data structure for most expensive rectilinear insertion exhibit a jump in
runtime by a factor of 15, much higher than the factor of 8 indicated by
the asymptotic analysis.  We believe that this jump is due to
exceeding the limits of the 32Kbyte level I cache on the 603e processor;
other jumps can be attributed to exceeding the Powerbook 3400's
256K level II cache.

The fact that the original Conga Line data structure
is slower than MultiConga and FastPair for rectilinear greedy maximum
matching with moderate input sizes, even though it performs fewer
distance computations, may be due to its higher space
usage causing poor caching behavior; note that the other two
methods become slower than it only on problem sizes at which they too
exceed the cache. 
Perhaps the relatively poor performance of the
quadtree method is also due to its high memory usage.  It would be of
interest to develop more cache-efficient closest pair data structures which
take better advantage of modern computer memory systems.

\bibliography{greedy}

\end{document}